\documentclass[11pt,twoside]{article}
\usepackage{asp2004}
\usepackage{psfig}
\usepackage{epsf}
\usepackage{epsfig}
\usepackage{wrapfig}
\usepackage{graphics}
\usepackage{lscape}
\def\edcomment#1{\iffalse\marginpar{\raggedright\sl#1\/}\else\relax\fi}
\reversemarginpar

\def\araa{{\em ARAA}}

\def\apj{{\em ApJ}}
\def\apjs{{\em ApJS}}
\def\aap{{\em A\&A}}

\def\mnras{{\em MNRAS}}

\def\prl{{\em PRL}}

\begin{document}
\title{Gravoturbulent Star Cluster Formation}
\author{Ralf S.\ Klessen\altaffilmark{1}, Javier Ballesteros-Paredes\altaffilmark{2}, Yuexing Li\altaffilmark{3,4},
  Mordecai-Mark Mac~Low\altaffilmark{3,4}}
\affil{$^{1}$Emmy Noether Research Group, Astrophysikalisches Institut Potsdam, An der Sternwarte 16,
 14482 Potsdam, Germany (rklessen@aip.de)
}
\affil{$^{2}$Centro de Radioastronom{\'\i}a y Astrof{\'\i}sica,
UNAM. Apdo. Postal 72-3 (Xangari), Morelia, Michoc{\'a}n 58089,
M{\'e}xico 
}
\affil{$^{3}$Department of Astronomy, Columbia University, New York,
NY 10027, USA 
}
\affil{$^{4}$Department of Astrophysics, American Museum of Natural
History, 79th Street at Central Park West, New York, NY 10024-5192,
USA 
}

\begin{abstract}
  
Stars form by gravoturbulent fragmentation of interstellar gas clouds.
The supersonic turbulence ubiquitously observed in Galactic  molecular gas
generates strong density fluctuations with gravity taking over in the
densest and most massive regions. Collapse sets in to build up stars
and star clusters.  

Turbulence plays a dual role. On global scales it provides support,
while at the same time it can promote local collapse.  Stellar birth
is thus intimately linked to the dynamical behavior of parental gas
cloud, which determines when and where protostellar cores form, and
how they contract and grow in mass via accretion from the surrounding
cloud material to build up stars.  Slow, inefficient, isolated star
formation is a hallmark of turbulent support, whereas fast, efficient,
clustered star formation occurs in its absence.

The fact that Galactic molecular clouds are highly filamentary can be
explained by a combination of compressional flows and shear.  The
dynamical evolution of nascent star clusters is very complex. This
strongly influences the stellar mass spectrum. The equation
of state (EOS) plays a pivotal role in the fragmentation process.
Under typical cloud conditions, massive stars form as part of dense
clusters.  However, for gas with effective polytropic index greater
than unity star formation becomes biased towards isolated massive
stars, which may be of relevance for understanding Pop III stars.
\end{abstract}

\section{Introduction}
\label{sec:intro}
Star clusters form by gravoturbulent fragmentation in interstellar
clouds. The supersonic turbulence ubiquitously observed in Galactic
gas clouds generates strong density fluctuations with gravity taking
over in the densest and most massive regions.  Once such cloud regions
become gravitationally unstable, collapse sets in and leads to the
formation of stars and star clusters. Yet the conditions for
fragmentation and the physical processes that govern the early
evolution of nascent star clusters are poorly understood.

Following up on analytical studies (starting with Jeans 1902; and
later  Larson 1969; Shu 1977;
Elmegreen 1993; Padoan 1995; Padoan \& Nordlund 2002), most current
investigations concentrate on a numerical approach to star cluster
formation.  For example, the effects of interstellar turbulence have
been studied extensively in a series of 3D simulations by Klessen,
Burkert, \& Bate (1998), Klessen, Heitsch, \& Mac Low (2000), Klessen
\& Burkert (2000, 2001), Heitsch, Mac Low, \& Klessen (2001a,b), Klessen
(2001). See also Ballesteros-Paredes et al.\ (1999ab, 2003), Padoan \&
Nordlund (1999), Padoan et al.\ (2001), Bate, Bonnell, \& Bromm
(2003) or Bonnell, Bate, \& Vine (2003).  A complete
overview is given in the reviews by Larson (2003) and Mac Low \& Klessen
(2004).

In this proceedings paper we call your attention to the dynamical
complexity arising from the interplay between supersonic turbulence
and self-gravity, and introduce the concept of gravoturbulent
fragmentation.  We argue that in typical star forming clouds
turbulence generates the density structure in the first place and then
gravity takes over in the densest and most massive regions to build up
the star cluster.  In Section 2 we focus on spatial distribution and
timescale of star formation, then in Section 3, we discuss a specific
example of a star forming filament similar to those observed in
Taurus, and in Section 4 we speculate about the mass spectra of clumps
and stars in the context of the gravoturbulent fragmentation model.
Finally, in Section 5 we demonstrate that the equation of state (EOS)
of the interstellar gas plays a pivotal role in gravoturbulent
fragmentation. The EOS determines whether molecular cloud regions
build up clusters of low to intermediate-mass stars, or form isolated
high-mass objects.

\section{Spatial Distribution and Timescale of Star Formation}
\label{sec:location-time}
Supersonic turbulence plays a dual role in star formation. While it
usually is strong enough to counterbalance gravity on global scales it
will usually provoke collapse locally (Mac~Low \& Klessen 2004).  Turbulence establishes a complex
network of interacting shocks, where regions of high-density build up
at the stagnation points of convergent flows.  These gas clumps can be
dense and massive enough to become gravitationally unstable and
collapse when the local Jeans length becomes smaller than the size of
the fluctuation.  However, the fluctuations in turbulent velocity
fields are highly transient.  They can disperse again once the
converging flow fades away (V{\'a}zquez-Semadeni, Shadmehri, \&
Ballesteros-Paredes 2002).  Even clumps that are strongly dominated by
gravity may get disrupted by the passage of a new shock front (Mac Low et al.\ 1994).

For local collapse to result in the formation of stars, Jeans unstable,
shock-generated, density fluctuations therefore must collapse to sufficiently
high densities on time scales shorter than the typical time interval between
two successive shock passages.  Only then do they `decouple' from the ambient
flow pattern and survive subsequent shock interactions.  The shorter the time
between shock passages, the less likely these fluctuations are to survive. The
overall efficiency of star formation depends strongly on the wavelength and
strength of the driving source (Klessen et al.\ 2000, Heitsch et al.\
2001). Both regulate the amount of gas available for collapse on
the sonic scale where turbulence turns from supersonic to subsonic
(V\'azquez-Semadeni, Ballesteros-Paredes, \& Klessen 2003).

The velocity field of long-wavelength turbulence is dominated by
large-scale shocks which are very efficient in sweeping up molecular
cloud material, thus creating massive coherent structures. These
exceed the critical mass for gravitational collapse by far, because
the velocity dispersion within the shock compressed region is much
smaller than in the ambient turbulent flow. The situation is similar
to localized turbulent decay, and quickly a cluster of protostellar
cores builds up. Both decaying and large-scale turbulence lead to a
{\em clustered} mode of star formation. Prominent examples are the
Trapezium Cluster in Orion with a few thousand young stars, but also
the Taurus star forming region which is historically considered as a
case of isolated stellar birth. Its stars have formed almost
simultaneously within several coherent filaments which apparently are
created by external compression (see Ballesteros-Paredes et al.\ 
1999a). This renders it a clustered star forming region in the sense
of the above definition.

The efficiency of turbulent fragmentation is reduced if the driving
wavelength decreases. There is less mass at the sonic scale and the
network of interacting shocks is very tightly knit. Protostellar cores
form independently of each other at random locations throughout the
cloud and at random times. There are no coherent structures with
multiple Jeans masses. Individual shock generated clumps are of low
mass and the time interval between two shock passages through the same
point in space is small.  Hence, collapsing cores are easily destroyed
again. Altogether star formation is inefficient. This scenario then
corresponds to an {\em isolated} mode of star formation. Stars that
truly form in isolation are, however, very rarely observed -- most
young stars are observed in clusters or at most loose aggregates. From
a theoretical point of view, there is no fundamental dichotomy between
these two modes of star formation, they rather define the extreme ends
in the continuous spectrum of the properties of turbulent molecular
cloud fragmentation.

Altogether, we call this intricate interaction between turbulence on the one
side and gravity on the other -- which eventually leads to the transformation of
some fraction of molecular cloud material into stars as described above -- {\em
  gravoturbulent fragmentation}. To give an example, we discuss in detail the
gravitational fragmentation in a shock-produced filaments that closely
resembles structures observed in the Taurus star forming region.

\section{Gravitational Fragmentation of a Filament in a Turbulent Flow}\label{Taurus:sec} 

In Taurus, large-scale turbulence is thought to be responsible for the
formation of a strongly filamentary structure (e.g.\
Ballesteros-Paredes et al.\ 1999a). Gravity within the filaments should then be
considered as  the main mechanism for forming cores and stars. Following earlier ideas
by Larson (1985), Hartmann (2002) has shown that the Jeans length within a
filament, and the timescale for it to fragment are given by
\begin{eqnarray}
\lambda_J  &= & 1.5 \ T_{10}\ A_V^{-1} \ {\rm pc,} \label{lambda-lee:eq} \\
\tau  & \sim  & \ 3.7 \ T^{1/2}_{10}\ A_V^{-1} \ {\rm Myr.}
        \label{timescale-filament:eq} 
\end{eqnarray}
where $T_{10}$ is the temperature in units of 10$\,$K, and $A_V$ is the
visual extinction through the center of the filament. 
By using a mean visual extinction for starless cores of $A_V\sim 5$, equation 1 gives a characteristic
Jeans length of $\lambda_J\sim 0.3\,$pc, and collapse should occur in
about 0.74$\,$Myr. Indeed, Hartmann (2002) finds $3-4$
young stellar objects per parsec with agrees well with the above
numbers from linear
theory of gravitational fragmentation of filaments.

In order to test these ideas, we resort to numerical simulations. We analyze a
SPH calculation (Benz 1990, Monaghan 1992) of a star forming region that was
specifically geared to the Taurus cloud. Details on the numerical
implementation, on performance and convergence properties of the method, and
tests against analytic models and other numerical schemes in the context of
turbulent supersonic astrophysical flows can be found in Mac~Low et al.\ 
(1998), Klessen \& Burkert (2000, 2001) and Klessen et al.\ (2000).

This simulation has been performed without gravity until a particular, well
defined elongated structure is formed. We then turn-on self-gravity. This
leads to localized collapse and a sparse cluster of protostellar cores builds
up. Timescale and spatial distribution are in good agreement with the Hartmann
(2002) findings in Taurus. For illustration, we show eight column density
frames of the simulation in Figure 1. The first frame shows
the structure just before self-gravity is turned-on, and we note that the
filament forms cores in a fraction of Myr. The timestep between frames is
0.1$\,$Myr. The mean surface density for the filament is
0.033$\,$g$\,$cm$^{-2}$, corresponding to a visual extinction of $\sim$7.5.
Using equations 1 and 2 this value
gives a Jeans length of $\lambda_J \sim 0.2\,$pc, and a collapsing timescale
of $\tau \sim 0.5\,$Myr. Note from Figure 1 that the first
cores appear roughly at $\tau \sim 0.3\,$Myr, although the final structure of
collapsed objects is clearly defined at $t=0.5\,$Myr. The typical
separation between protostellar cores (black dots in Figure 1) is about the Jeans length
$\lambda_J$.

\begin{figure}
\centerline{\psfig{figure=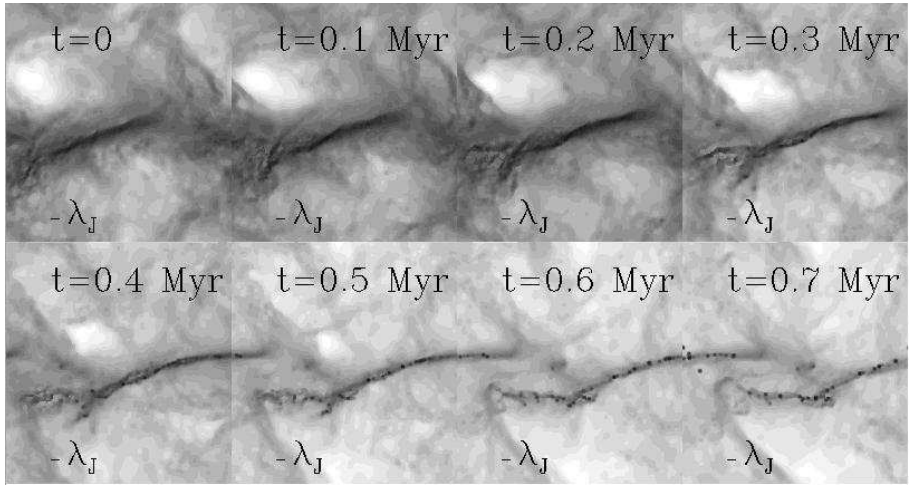,width=120truemm, angle=0}}
\caption{%
  Evolution of the column density of an SPH simulation. The filament
  in the first frame (before self-gravity is turned-on) shows that
  turbulence is responsible in forming this kind of structures. The
  small bar in the bottom-left of each frame denotes the Jeans length
  (equation 1) at this time. At later times, self-gravity is turned on
  and the filament suffers gravitational fragmentation on a free-fall
  timescale (equation 2).}
\end{figure}

This example demonstrates that indeed turbulence is able to produce a strongly
filamentary structure and that at some point gravity takes over to form
collapsing objects, the protostars. However, the situation is quite complex.
Just like in Taurus, the filament in Figure 1 is not a perfect cylinder, the
collapsed objects are not perfectly equally spaced as predicted by idealized
theory, and protostars do not form simultaneously but during a range of times
(between $t \approx 0.3$ and 0.6$\,$Myr). Even though the theory of
gravitational fragmentation of a cylinder appears roughly, it becomes clear
that the properties of the star forming region not only depend on the
conditions set initially but are influenced by the large-scale turbulent flow
during the entire evolution. Gravoturbulent fragmentation is a continuous
process that shapes the accretion history of each protostar in a stochastic
manner (e.g.\ Klessen 2001a).

\begin{figure}[t]
\unitlength1cm
\begin{center}
\begin{picture}(10.0,8.0)
\put( 0.0, -0.8) {\epsfxsize=10.0cm \epsfbox{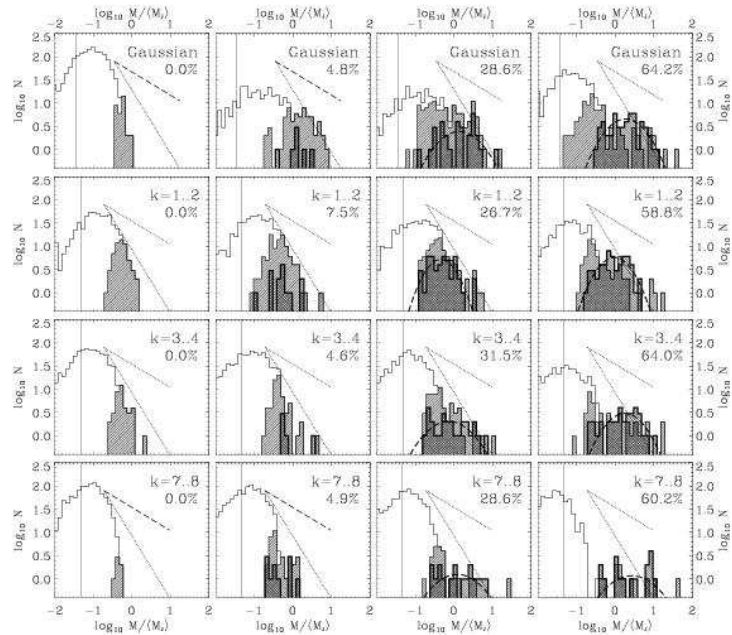}}
\end{picture}
\end{center}
\caption{Mass spectra of protostars (hatched thick-lined
  histograms), of gas clumps (thin lines), and of the subset of Jeans
  unstable clumps (thin lines, hatched distribution). Different 
  evolutionary phases are defined  by the fraction of mass converted into
  protostars and are indicated in the upper right corner of each plot. 
  Masses are
  binned logarithmically and normalized to the average Jeans mass
  $\langle M_{\rm J}\rangle$. (From Klessen 2001b.)
}
\end{figure}

\section{Mass Spectra of Clumps and Protostellar Cores}
\label{sec:mass-spectra}
The dominant parameter determining stellar evolution is the mass. We
discuss now how the final stellar masses may depend on the
gravoturbulent fragmentation process, and analyze four numerical
models which span the full parameter range from strongly clustered to
very isolated star formation (for full detail see Klessen 2001b).

Figure 2 plots the mass distribution of all gas clumps, of the subset
of Jeans critical clumps, and of collapsed cores. We show four
different evolutionary phases, initially just when gravity is
`switched on', and after turbulent fragmentation has lead to the
accumulation of $M_{\rm \large *}\approx 5$\%, $M_{\rm \large
  *}\approx 30$\% and $M_{\rm \large *}\approx 60$\% of the total mass
in protostars.
In the  completely pre-stellar phase the clump mass spectrum
is  very steep (about Salpeter slope or less) at the high-mass end. It
has a break and
gets shallower below $M \approx 0.4 \,\langle M_{\rm J} \rangle$ with
slope $-1.5$. The spectrum strongly declines
beyond the SPH resolution limit. Individual clumps are
hardly more massive than a few $\langle M_{\rm J}\rangle$.
Gravitational evolution modifies the distribution
of clump masses considerably. As clumps merge and grow bigger, the
spectrum becomes flatter and extends towards larger masses.
Consequently the number of cores that exceed the Jeans limit
increases. This is most evident in the Gaussian model of decayed
turbulence, the clump mass spectrum exhibits a slope $-1.5$.


The mass spectrum depends on the wavelength of the dominant velocity
modes.  Small-scale turbulence does not allow for massive, coherent
and strongly selfgravitating structures. Together with the short
interval between shock passages, this prohibits efficient merging and
the build up of a large number of massive clumps. Only few clumps
become Jeans unstable and collapse to form stars. This occurs at
random locations and times. The clump mass spectrum remains steep.
Increasing the driving wavelength leads to more coherent and rapid
core formation, resulting in  a larger number of protostars.

Long-wavelength turbulence or turbulent decay produces a core mass
spectrum that is well approximated by a {\em log-normal}. It roughly
peaks at the {\em average thermal Jeans mass} $\langle M_{\rm
  J}\rangle$ of the system (see Klessen \& Burkert 2000, 2001) and is
comparable in width with the observed IMF (Kroupa 2002). 
The log-normal shape of the mass distribution may be
explained by invoking the central limit theorem (e.g.\ Zinnecker
1984), as protostellar cores form and evolve through a sequence of
highly stochastic events (resulting from supersonic turbulence and/or
competitive accretion).

\section{Effects of the Equation of State}
So far, we concentrated on isothermal models of Galactic molecular
clouds. More generally, however, the balance of heating and cooling in
a molecular cloud can be described by a polytropic EOS, $P =
K\rho^{\gamma}$, where $K$ is a constant, and $P, \rho$ and $\gamma$
are thermal pressure, gas density and polytropic index, respectively.
A detailed analysis by Spaans \& Silk (2000) suggests that $0.2 <
\gamma < 1.4$ in the interstellar medium.

\begin{figure}[tb]
\leavevmode
\includegraphics[height=1.2in]{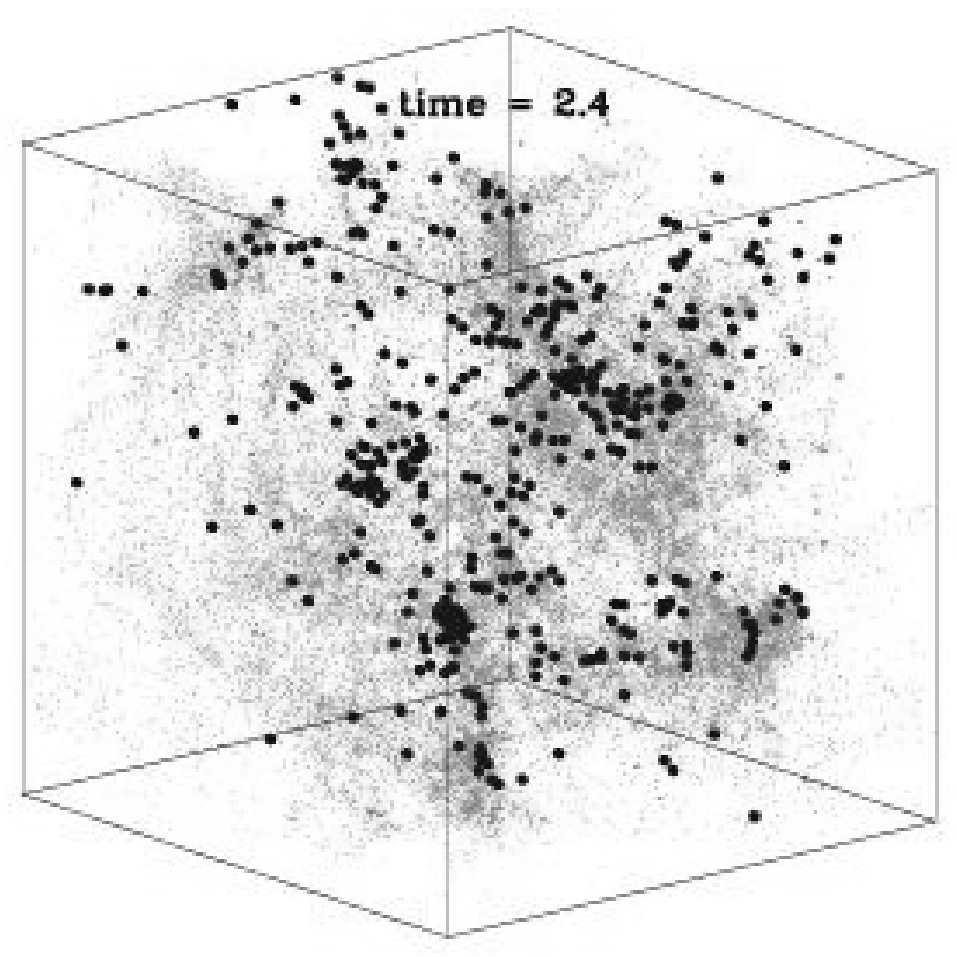}
\includegraphics[height=1.2in]{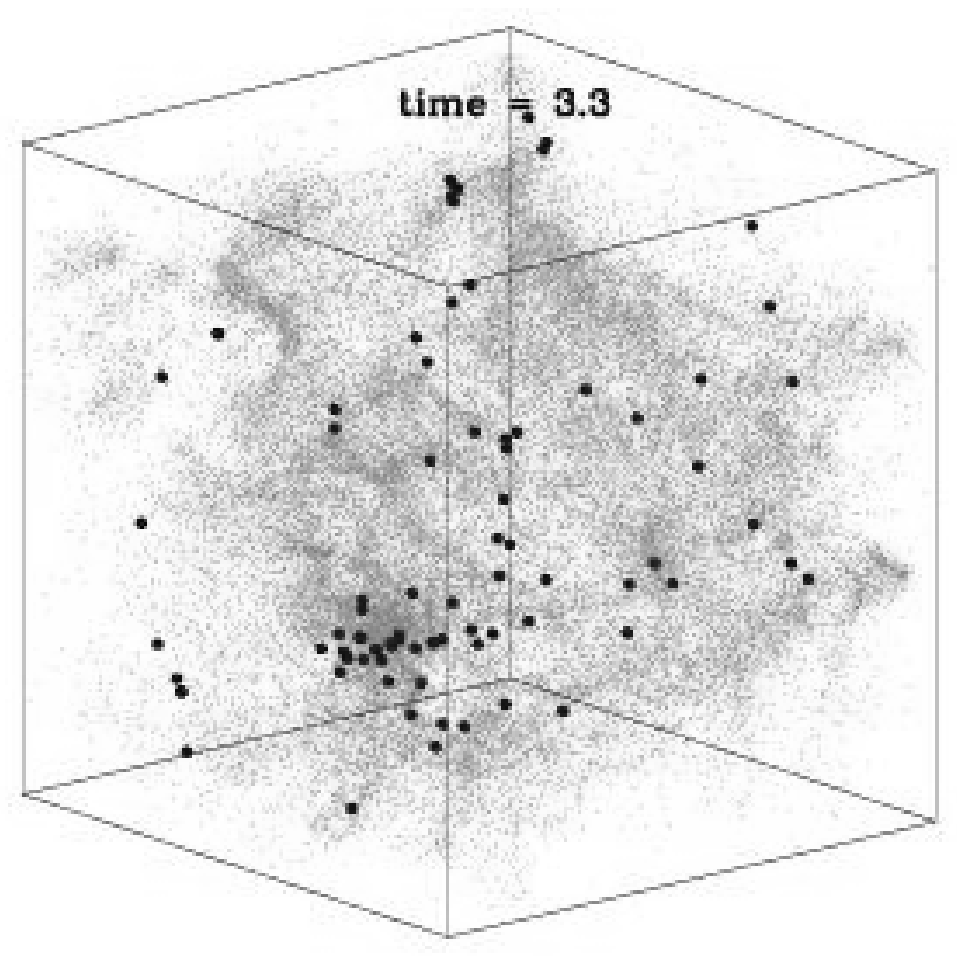}
\includegraphics[height=1.2in]{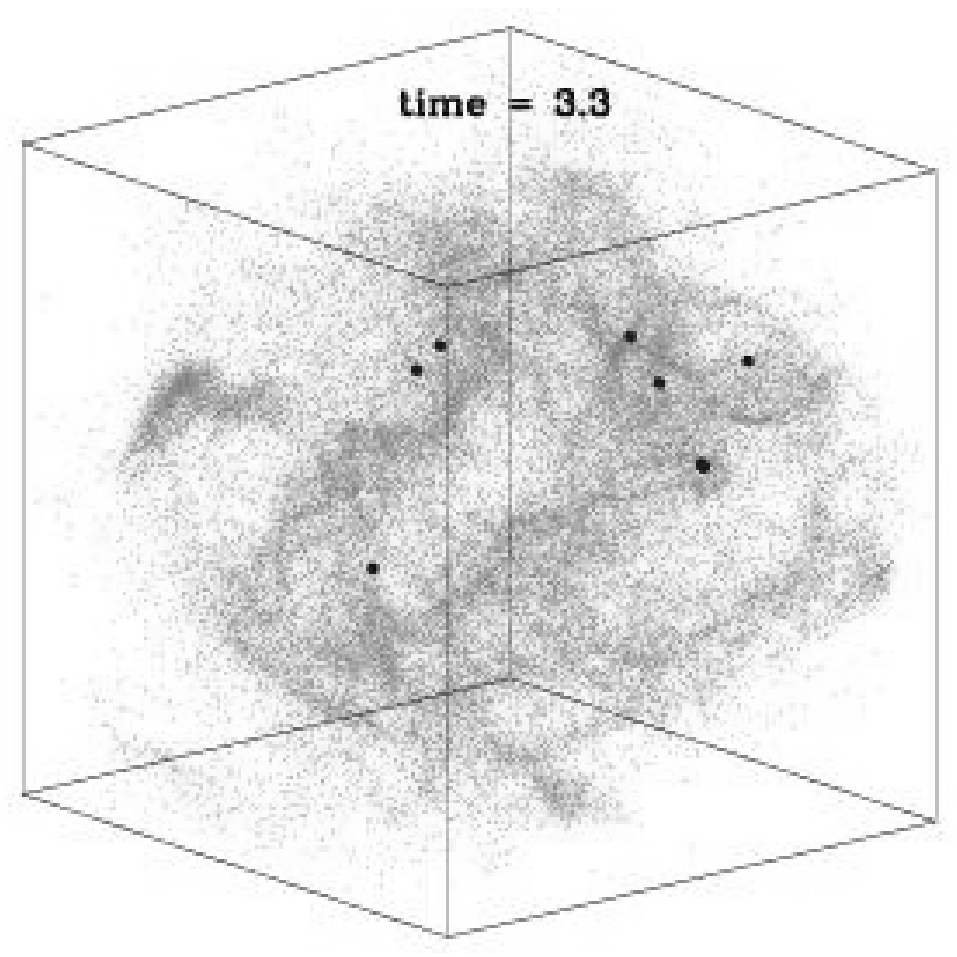}
\includegraphics[height=1.0in]{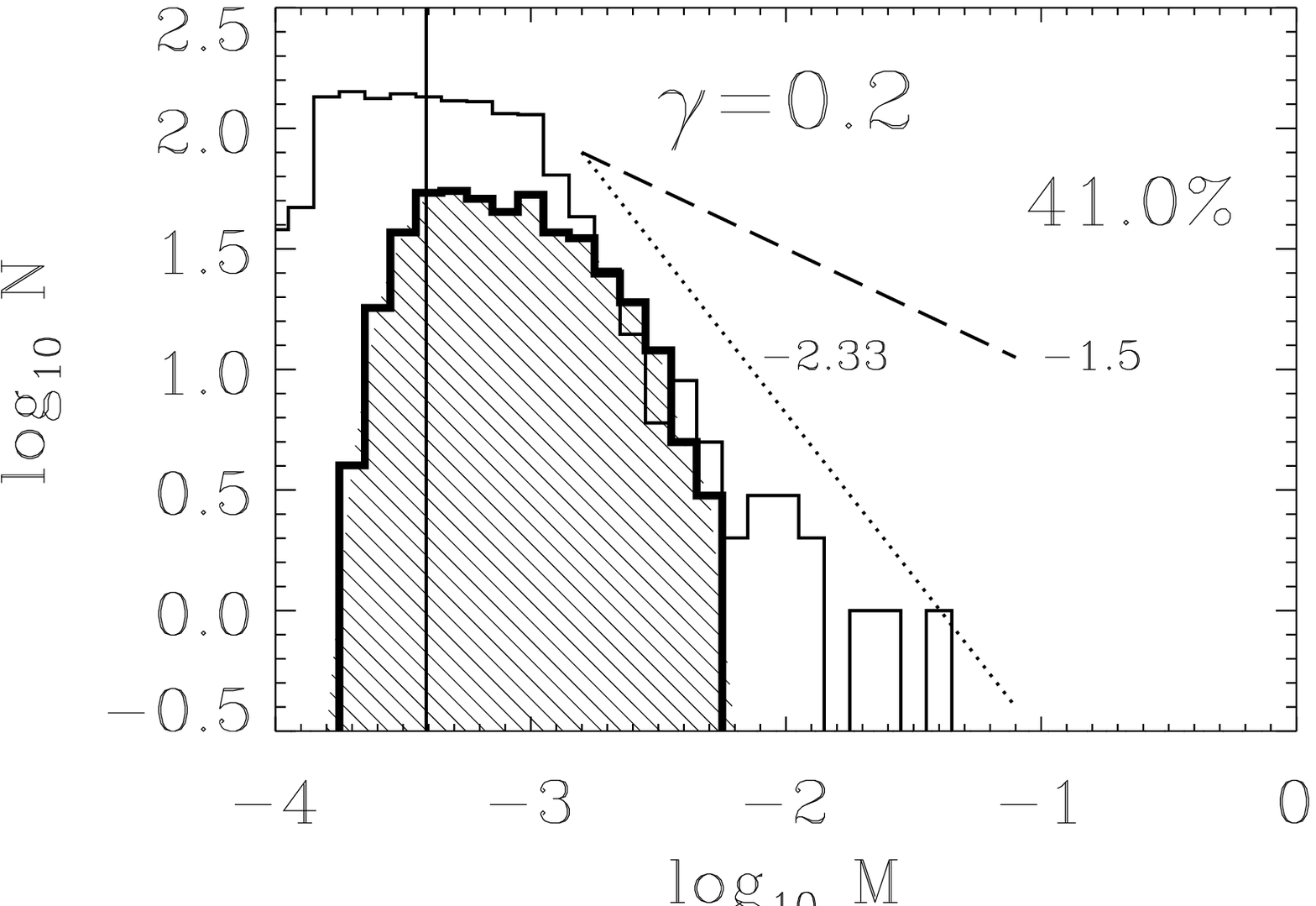}
\mbox{~~~~~~~~}
\includegraphics[height=1.0in]{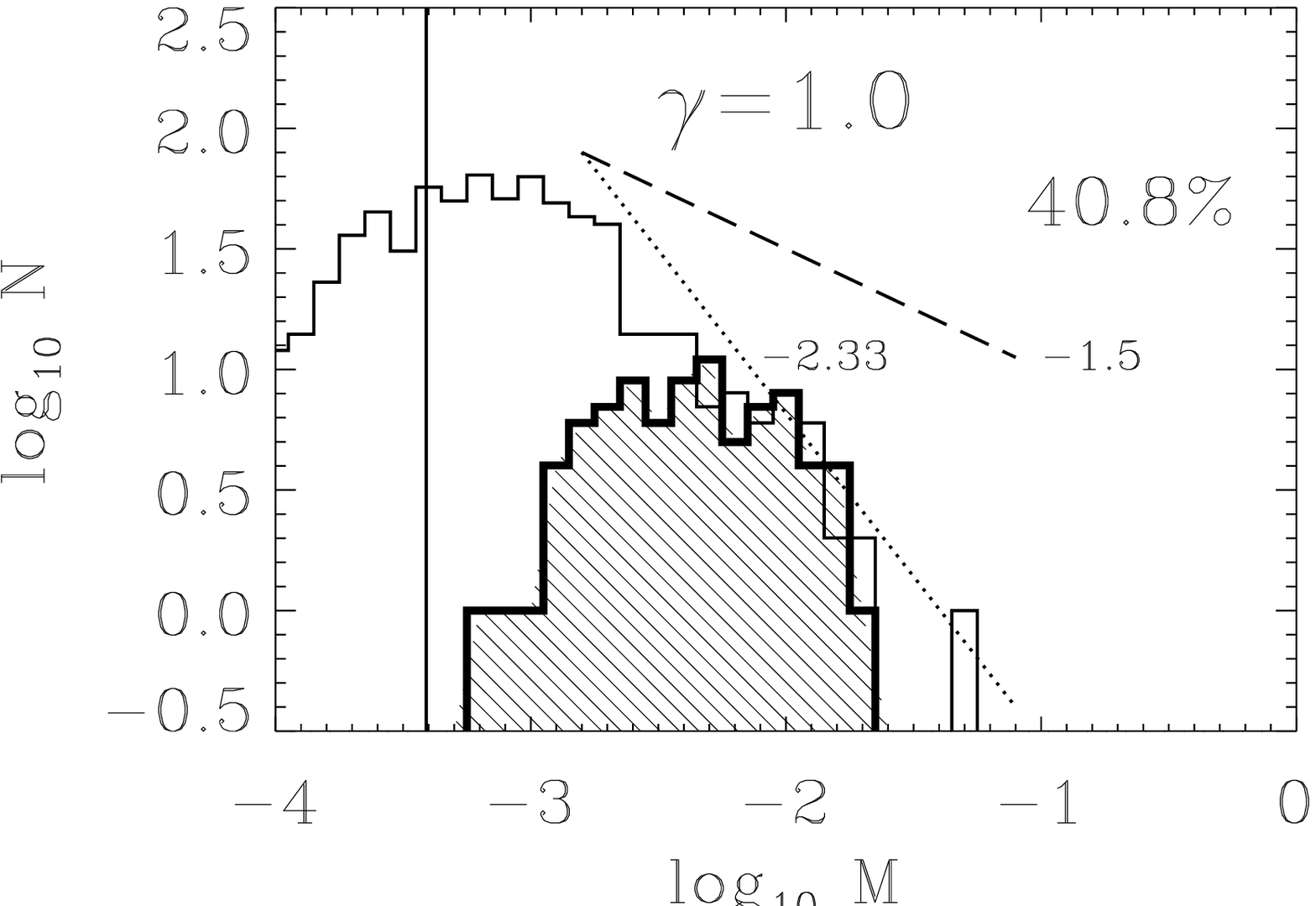}
\mbox{~~~~~~~~}
\includegraphics[height=1.0in]{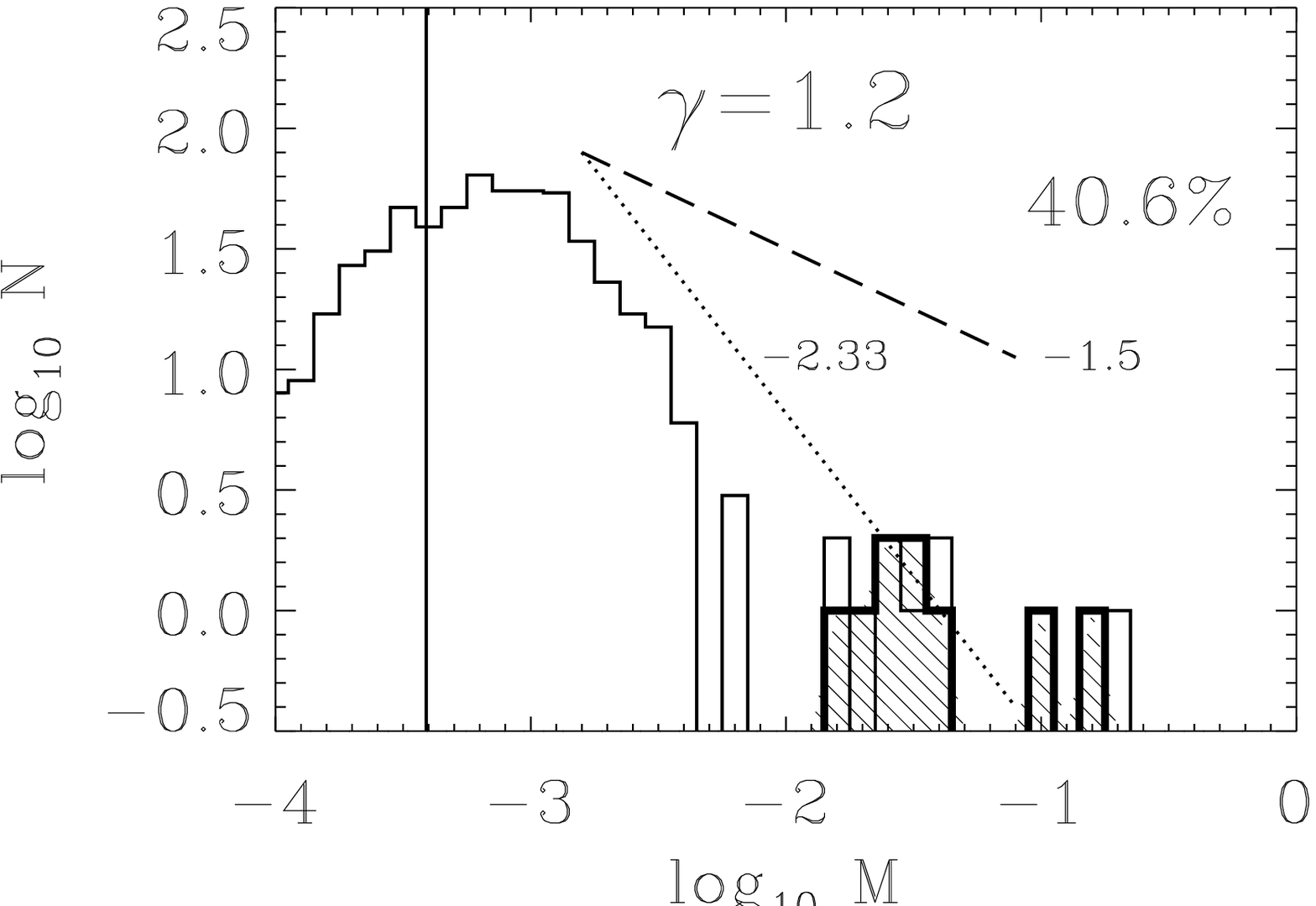}

\caption{\label{fig_mas}Top: 3-D distribution of the gas and protostars for
    different $\gamma$. Bottom: Mass spectra of gas clumps
    (\textit{thin  
    lines}) and of protostars (collapsed cores: \textit{hatched thick-lined
    histograms}) for the corresponding cube above. The
    percentage shows the fraction of total mass accreted onto protostars. The
    vertical line shows the SPH resolution limit. Shown also are two
    power-law spectra with $\nu = -1.5$ (dashed-line) and $\nu =
    -2.33$ (dotted line). (Figure adopted from Li et al.\ 2003.)}
\end{figure}

Li, Klessen \& Mac~Low (2003) carried out detailed smoothed particle
hydrodynamics (SPH) simulations to determine the effects of different
EOS on gravoturbulent fragmentation by varying $\gamma$ in steps of
0.1 in otherwise identical simulations. Figure 3 illustrates how low
$\gamma$ leads to the build-up of a dense cluster of stars, while high
values of $\gamma$ result in isolated star formation. It also
shows that the spectra of both the gas clumps and protostars change
with $\gamma$.  In low-$\gamma$ models, the mass distribution of the
collapsed protostellare cores at the high-mass end is roughly
log-normal.  As $\gamma$ increases, fewer but more massive cores
emerge. When $\gamma > 1.0$, the distribution is dominated by high
mass protostars only, and the spectrum tends to flatten out. It is no
longer described by either a log-normal or a power-law. The clump mass
spectra, on the other hand, do show power-law behavior at the high
mass side, even for $\gamma > 1.0$.

This suggest that stars tend to form in clusters in a low-$\gamma$
environment. Protostellar cores are of low mass in this case. The apparent lack of
power-law behavior for the cores in the protostellar cluster might
imply that simple accretion is unable to generate as many high-mass
stars as predicted by the observations, hinting that
other mechanisms such as collisions (Bonnell, Bate \& Zinnecker 1998)
may be at work to produce the massive stars in a cluster. Higher
resolution models will be necessary to confirm this, however.

On the other hand, our results also imply that massive stars can form
in small groups or alone in gas with $\gamma > 1.0$.
Spaans \& Silk (2000) suggest that a stiffer EOS ($\gamma > 1.0$) leads
to a peaked IMF, biased toward massive stars, while an EOS with
$\gamma < 1.0$ results in a power-law IMF, in general agreement with our
simulations.

The formation of isolated massive stars is of great interest, as usually, massive stars are found in clusters. But recently, Lamers
et al.\ (2002) reported observations of isolated massive
stars or very small groups of  massive stars in the bulge
of M51. Also Massey (2002) finds  massive, apparently isolated field stars
in both the Large and Small Magellanic Clouds. From our
simulations, we see that when $\gamma > 1$, only very few or possibly
only one fragment occurs. These then are  massive, and  would result in the
formation of  high-mass stars.

High resolution simulations by Abel, Bryan \& Norman (2002) of the
formation of the first star suggest that initially only one massive
metal-free star forms per pregalactic halo. In the early Universe,
inefficient cooling due to the lack of metals may result in high
$\gamma$.  Our models then suggest weak fragmentation, resulting
in the formation of only one massive star per cloud.


\acknowledgements{We thank for support from various sources: RSK from
  the Emmy Noether Program of the Deutsche Forschungsgemeinschaft
  (grant no.\ KL1358/1); JBP from Conacyt's grant I39318-E; MMML from
  a NASA ATP grant NAG5-10103, and a NSF CAREER grant AST99-85392.  }

\end{document}